# Socio-Physical Approach to Consensus Building and the Occurrence of Opinion Divisions Based on External Efficacy

Yasuko Kawahata [†]

Faculty of Sociology, Department of Media Sociology, Rikkyo University, 3-34-1 Nishi-Ikebukuro,Toshima-ku, Tokyo, 171-8501, JAPAN.
ykawahata@rikkyo.ac.jp

**Abstract:** The proliferation of public networks has enabled instantaneous and interactive communication that transcends temporal and spatial constraints. The vast amount of textual data on the Web has facilitated the study of quantitative analysis of public opinion, which could not be visualized before. In this paper, we propose a new theory of opinion dynamics. This theory is designed to explain consensus building and opinion splitting in opinion exchanges on social media such as Twitter. With the spread of public networks, immediate and interactive communication that transcends temporal and spatial constraints has become possible, and research is underway to quantitatively analyze the distribution of public opinion, which has not been visualized until now, using vast amounts of text data. In this paper, we propose a model based on the Like Bounded Confidence Model, which represents opinions as continuous quantities. However, the Bounded Confidence mModel assumes that people with different opinions move without regard to their opinions, rather than ignoring them. Furthermore, our theory modeled the phenomenon in such a way that it can incorporate and represent the effects of external external pressure and dependence on surrounding conditions. This paper is a revised version of a paper submitted in December 2018(Opinion Dynamics Theory for Analysis of Consensus Formation and Division of Opinion on the Internet).

## 1. Introduction

The study of opinion dynamics has a long history and has been the subject of much research, mainly in the field of sociology. Early studies assumed linearity, but models incorporating nonlinearity have also been studied. Consensus formation based on local majority rule has been studied as an application of renormalization group theory in physics. The two phases are separated by a threshold, and near the threshold the system becomes very sensitive to minute changes in parameters. The first model population stage in this opinion dynamics was one in which the system is developed as a region of parameters with two possible states: a "work state" and a "strike state"[16-18]. This approach was the catalyst for a variety of other approaches. These include the Ising model, in which opinions are discrete and resolve extreme conflict compositions; the Voter model, in which agents affect only one of their neighbors at a time; the Defiant model, in which opinions are continuous; and the Hegzelman-Krauss model[1-7], in which opinions are bounded. All of these models are based on bounded "confidence intervals," and a simulation model is proposed in which the rule for setting parameters is that agents are not influenced by people whose opinions differ significantly from their own.

In the preceding studies, we again touch on the evolution of the theory of opinion dynamics. The society we live in has undergone a major social transformation over the past 100 years, with the spread of public networks changing the forms of communication. Also, the theory of magnetic physics, which compares the agreement and disagreement of opinions with the direction of the magnetic moment of magnetism, has been studied in the field of social physics by Garam et al[16-18]. Many mathematical theories of opinion dynamics treat opinions as discrete values of +1 and 0 or +1 and -1. In contrast, some theories consider opinions as continuous values, which can be varied through the exchange of opinions with others. The bounded belief model is a representative model of theories that treat opinions as continuous transitions[8-15]. In this study, we propose a theory that represents opinions as continuous values and handles changes in opinion values through exchanges of opinions with others. Furthermore, we assume that each person's opinion can be either a positive or negative value. For example, some studies of tweets about the political situation in the U.S. classify political opinions on a one-dimensional axis from conservative to liberal. In this study, we assume, as in this literature, that differences in opinion can be expressed in terms of values on a unidimensional axis. Based on this theory, it is possible to represent the division of opinions in society by assuming that people with different opinions exchange their opinions and that their opinions are further divided. Such a division of opinions is a common phenomenon in social media such as X, bulletin boards[19-24].

# 2. Modelling opinion dynamics

Our model is based on the original bounded confidence model of Hegselmann-Krause Model. For a fixed agent, say $i$, where $1 \leqq i \leqq N$, we denote the agent's opinion at time $t$ by $I_i(t)$, Ishii-Kawahata(2018)[25].

$$i(t+1) = \sum_{j=1}^{N} D_{ij} I(t) \quad (1)$$

$$\Delta I_i(t) = c_i A(t) \Delta t + \sum_{j=1}^{N} D_{ij} I_j(t) \Delta t \quad (2)$$

$$Opinion_i(t) = \tanh(I_i(t)) \quad (3)$$

$$\Delta I_i(t) = c_i A(t) \Delta t + \sum_{j=1}^{N} D_{ij} I_j(t) \Delta t \quad (4)$$

## 2.1 Explanation of Parameters

$i(t+1)$: Value of $i$ at time $t+1$.

$I(t)$: Vector of overall values at time $t$.

$D_{ij}$: Element at the $i$th row and $j$th column of the matrix $D$. This represents the degree of interaction or influence.

$\Delta I_i(t)$: Amount of change in $I_i$ at time $t$.

$c_i$: $i$th element of the constant vector $c$. This represents the external influence received by individual entities.

$A(t)$: Sum of $I(t)$ at time $t$, indicating the overall "activity" or "intensity" of opinions.

$Opinion_i(t)$: Opinion or stance of $i$ at time $t$. The tanh function returns results in the range of -1 to 1, representing a spectrum from strongly negative opinions to strongly positive ones.

## 2.2 Interpretation in Social Phenomena

These equations appear to model interactions or influences between entities. Specifically, the opinion or stance $Opinion_i(t)$ of an entity $i$ changes based on the external influence $c_i A(t)$ it receives and the direct influence $D_{ij} I_j(t)$ from other entities. The matrix $D$ might represent the strength and direction of interactions between entities. For instance, it could be suited for modeling the influence from friends in a social network or the process of diffusion and acceptance of information.

## 2.3 Opinion dynamics for two agents

Let us first consider the case where the opinions of the two agents are the same. In this case, both opinions are positive. If $D_{ij} > 0$, $D_{ij}$ and $I_j(t)$ is positive. Thus, the opinion $I_i(t)$ moves in the positive direction as shown in fig.**??**. This means that by having a conversation with an agent of the same positive opinion, agent $i$ will change its opinion to be more and more positive. Similarly, if the opinions of both agents are the same negative opinion, the opinions become more and more negative.

## 2.4 Equation fot two agents

$$\Delta I_i(t) = -\alpha I_i(t) + c_i A(t) \Delta t + \sum_{j=1}^{N} D_{ij} I_j(t) \Delta t \quad (5)$$

### 2.4.1 Parameter Description

$\Delta I_i(t)$: Change in $I_i$ at time $t$.

$\alpha$: A constant representing the rate at which individual opinions or stances naturally decay over time.

$I_i(t)$: Intensity of opinion or stance of entity $i$ at time $t$.

$c_i$: $i$-th element of a constant vector representing the influence entity $i$ receives from the outside.

$A(t)$: Sum of $I(t)$ at time $t$, indicating the overall 'activity' or 'intensity' of the opinions or stances.

$D_{ij}$: Element of matrix $D$ at row $i$ and column $j$, indicating the degree of interaction or influence between entities.

$N$: Total number of entities.

**Explanation**

Equation (5) models the temporal variation of the opinion or stance of entity $i$. The opinion or stance $I_i(t)$ of each entity is influenced by three main factors:

$-\alpha I_i(t)$: This term represents how opinions or stances naturally decay over time. It could model, for example, how the freshness of information or interest decreases over time.

$c_i A(t) \Delta t$: This term represents the influence from the outside. The intensity of the influence that the specific entity $i$ receives is modulated by $c_i$. The overall activity or intensity $A(t)$ is based on the sum of opinions or stances of all entities.

$\sum_{j=1}^{N} D_{ij} I_j(t) \Delta t$: This term represents the direct influence from other entities. Each element $D_{ij}$ of the matrix $D$ indicates the strength of interaction or influence from entity $j$ to entity $i$.

This model illustrates how interactions between entities and influences from the outside impact the collective evolution of opinions or stances.

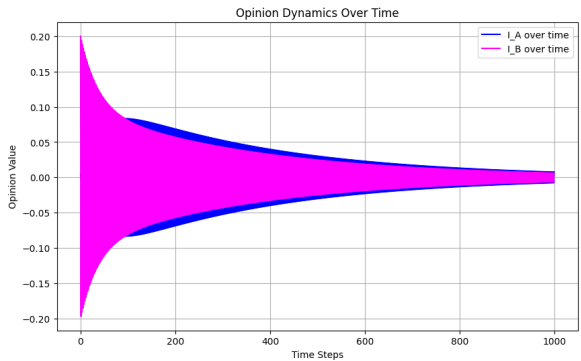

Fig. 1: Calculation result for N=2

Figure 1, dynamics of opinions over time for two agents, labeled as $I_A$ and $I_B$. The opinions of these agents change based on a combination of internal dynamics and the influence from the other agent.

**Figure 1 of Parameters**

Number of agents: $N = 2$

Interaction matrix $D$: This represents the extent to which each agent affects the other. Here, $D_{AB} = 1.0$ signifies the influence of agent B on agent A, while $D_{BA} = 0.5$ represents the influence of agent A on agent B.

Opinion constants $c$: Both agents have a constant value of 0.5, represented as $c = [0.5, 0.5]$.

The parameter $\alpha$: Set to 1.99, it determines the rate of opinion change.

Time steps: The simulation runs for 1000 time steps, with a time step interval $dt = 0.01$.

Initial opinions: Agent A starts with an opinion value of $I_A(0) = 0.005$ while Agent B has an initial value of $I_B(0) = 0.2$.

The simulation computes the change in opinion values for each agent at every time step using the formula:

$$\Delta I = -\alpha \times I(t) + c \times A(t) \times dt + D \times I(t) \times dt \quad (2)$$

where $A(t)$ is the sum of the opinions at time $t$. This delta is then added to the current opinion value to get the opinion value for the next time step.

The resulting plot, titled "Opinion Dynamics Over Time", visualizes the evolution of opinions for both agents over the 1000 time steps. The x-axis represents the time steps, and the y-axis represents the opinion value. The blue curve indicates the opinion of agent $I_A$ over time, and the magenta curve represents the opinion of agent $I_B$.

**Simulation Explanation**

Figure 1 is intended to simulate a specific phenomenon using a mathematical model. Specifically, it models the evolution of two variables (or states) with respect to time and draws the behavior of these states with respect to various "alpha" parameter values, resulting in the following functions for each parameter and formula.

**Parameters**

**timesteps, dt:**

**N:** This indicates the number of states in the system. In this case, it means that there are two different states or variables.

**D:** This can be thought of as an "interaction matrix" and shows how each state affects the other. In this case, there is asymmetry in the way one state affects the other because D has non-diagonal elements.

**c:** This array is a constant that controls the external input (or influence) in each state. It provides a constant "drive" or "push" to the system.

**timesteps, dt:** these variables control the time frame settings for the simulation. *timesteps* indicates the total number of steps in the simulation, and *dt* indicates the length of each time step.

**alphas:** This is the main parameter that varies in the simulation. Different alphas values test how the dynamics of the system change. This corresponds to changing the "sensitivity" or "response strength" of the system.

**Analysis based on the Opinion Dynamics Model(From Fig.1 : Case Study)**

**1. Analysis of the consensus formation process between the two parties:**

From the graph, one can observe the dynamic changes in the opinions of the two parties. Initially, the opinion of $I_B$ is high while that of $I_A$ is low. As time progresses, the opinion of $I_B$ decreases, and that of $I_A$ increases. Eventually, both opinions seem to converge to nearly the same value. This behavior indicates that both parties are taking into account each other's opinions and moving towards an agreement. Specifically, the values in D ($D_{AB} = 1.0$ and $D_{BA} = 0.5$) suggest that $I_A$ is more susceptible to the opinion of $I_B$.

**2. Cases and opinion formation in society:**

This model can be interpreted as mimicking the process of opinion convergence or consensus formation between individuals or groups in society. For instance, when one group is more influential than the other, or when information from a certain source has a strong influence on one group. The spread and influence of information through mass media or social media can also be represented using such a model.

### 3. Analysis based on the value of $\alpha$:

Alpha ($\alpha = 1.99$) is the parameter that determines the rate of change of opinions. The larger the value of $\alpha$, the faster the opinions change. Specifically, when $\alpha$ is large, there's a higher likelihood for the opinions of each agent to converge or diverge rapidly. Conversely, when $\alpha$ is small, the change in opinions tends to be more gradual. In this simulation, with $\alpha$ having a relatively high value of 1.99, one can observe that the convergence of opinions is progressing rapidly.

**Function**

> **Calculate delta_I:** This calculates the change in state at each time step. This change is based on the current state (I[:, t]), the interaction between states (np.dot(D, I[:, t])), external inputs (c * A_t), and the response strength of the system (-alpha * I[:, t]).

**Interpretation as a Social Phenomenon**

> The D matrix may represent the interaction between different social factors or groups. For example, how one group's opinion affects another group.
>
> *c* represents the influence of external stimuli, which may correspond to external factors that influence individual opinions and behaviors, such as media, advertising, government policies, etc.
>
> $\alpha$ can be interpreted as a parameter indicating how well a member of a society resists or responds to external pressures. A high $\alpha$ value indicates that the society is very sensitive to external influences, while a low $\alpha$ indicates that the society is more resistant to these influences.

Visualizing how the behavior of the system changes in each scenario provides a hypothesis for understanding the dynamics of collective behavior under various social pressures and external influences. This type of model could be applied to mimic diverse social phenomena such as opinion formation, cultural change, and information dissemination. It is also important to consider that a detailed examination of the specific sociological and psychological theories behind the model is necessary to understand exactly how each parameter corresponds to a particular factor or influence in the real world.

In Fig.2, Analysis from the Code and Provided Graph

### 1. Consensus formation process between two parties:

In the initial values of $\alpha$ (from 0.1 to 1.82), both lines (blue and magenta) seem to converge towards the same opinion value. This indicates a consensus of opinion has been established between the two parties. However, around $\alpha = 1.88$, a different pattern appears in the graph where both opinions fluctuate without fully converging.

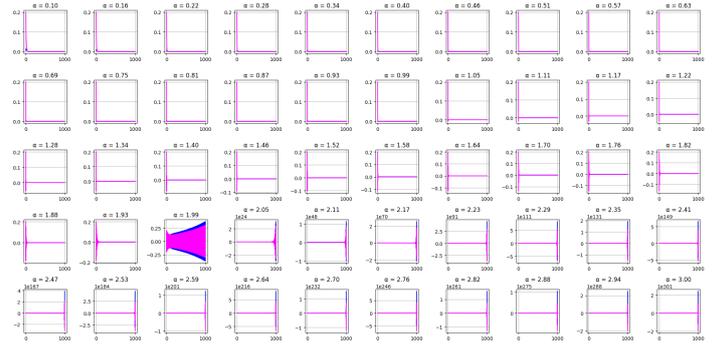

Fig. 2: Calculation result for N=2 by timesteps

### 2. Cases and opinion formation in a society:

This model could potentially mimic the propagation of information and the formation of opinions between groups. In the case of lower $\alpha$ values, the information spreads smoothly, indicating that people's opinions reach a consensus. However, as the value of $\alpha$ increases, the propagation of information becomes unstable, suggesting difficulties in reaching a consensus.

### 3. Analysis from step-by-step $\alpha$ values:

As the value of $\alpha$ increases, the rate of change in opinions also seems to increase. When $\alpha$ exceeds 1.88, the opinion fluctuations become significant, making it difficult to form a consensus. Especially when $\alpha$ exceeds 2, the opinion clashes intensify and it does not seem to reach a stable consensus.

### 4. Tendency of opinion clash based on $\alpha$ values:

Once $\alpha$ exceeds 1.88, there is no observed convergence in opinions and there are intense fluctuations and clashes in opinions. This possibly suggests that once $\alpha$ surpasses a certain threshold, opinion formation becomes unstable, making consensus formation difficult. Specifically, the fluctuations in opinions can be most vividly seen around an $\alpha$ value near 2.0.

## 3. Opinion dynamics for three agents

Next, calculations in the case of three people are shown. A has a positive opinion, B has a negative opinion, and a third person C has an almost neutral opinion.

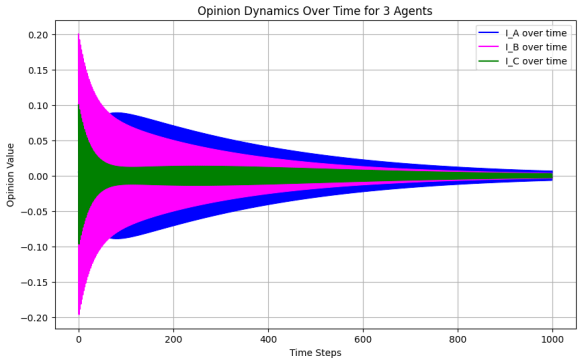

Fig. 3: Calculation result for N=3

In a situation similar to Fig.3, C is almost neutral in opinion, not much trusted by A and B, and $D_{AC}$ and $D_{BC}$ are small values. The calculation result is shown in Fig.3.

**Parameters and Initial Conditions:**

Number of agents, $N$: 3

Interaction matrix, $D$: $\begin{bmatrix} 0 & 1.0 & 0.5 \\ 0.5 & 0 & 0.5 \\ 0.5 & 0.5 & 0 \end{bmatrix}$

Common parameter for all agents, $c$: [0.5, 0.5, 0.5]

$\alpha$ value: 1.99

Total time steps: 1000 with a time difference, $\Delta t$: 0.01

Initial opinion values for the 3 agents: 0.005, 0.2, and 0.1 respectively.

**Simulation Process:**

The opinion value for each agent is updated at each time step based on the given interaction matrix, $D$, and the common parameter, $c$. The term $-\alpha \times I_{i,t}$ represents the influence of the opinion value of agent $i$ at time $t$. The term $c \times A_t \times \Delta t$ represents the common influence factor for all agents, where $A_t$ is the sum of opinion values of all agents at time $t$. The interaction matrix $D$ captures the influence of other agents on a given agent's opinion.

**Resultant Graph Analysis:**

The graph titled "Opinion Dynamics Over Time for 3 Agents" plots the opinion values of the 3 agents over 1000 time steps. The agents are labeled as I_A (blue), I_B (magenta), and I_C (green). From the graph, it's evident that:

Agent I_A starts with a low initial opinion value but quickly converges towards a stable opinion.

Agent I_B starts with a high initial opinion value and converges towards a lower stable opinion over time.

Agent I_C has an intermediate initial opinion value and stabilizes around a value close to that of Agent I_A.

The interactions between these agents, as defined by the interaction matrix $D$, play a crucial role in shaping their opinion dynamics over time.

**Analysis of Opinion Dynamics for 3 Agents**

**(1) Examination of the consensus formation process among the three agents:**

Initial opinion values are set as $I_A$ : 0.005, $I_B$ : 0.2, and $I_C$ : 0.1.

As the simulation progresses, the opinion values of the three agents fluctuate while influencing each other. It appears that the opinion values eventually converge.

**(2) Possible real-world scenarios and opinion formation:**

This model represents how opinions change through exchanges and interactions. Specifically, it may relate to scenarios like social networking, discussions within a group, and opinion formation.

**(3) Insights based on the value of $\alpha$ at each step:**

$\alpha$ is set to 1.99, representing the self-decay rate of the agent's opinion. A higher value of $\alpha$ suggests that the agent's opinion rapidly diverges from its previous stance.

With a high value of 1.99 for $\alpha$, agents tend to maintain their distinct opinions, but can also change them rapidly due to the influence of other agents.

**(4) The correlation between the value of $\alpha$ and the tendency of opinion clashes among the three agents:**

A high $\alpha$ value implies that agents hold their opinions strongly and are prone to rapid changes when influenced by other agents. This can lead to a higher potential for opinion clashes and significant fluctuations.

However, in this simulation, due to interactions between agents and external influences, opinions tend to converge eventually.

In summary, this simulation illustrates how interactions between agents and external influences can affect opinion dynamics. Additionally, the value of $\alpha$ can influence the variability and speed of opinion convergence.

# 4. Discussion

In Fig.4,

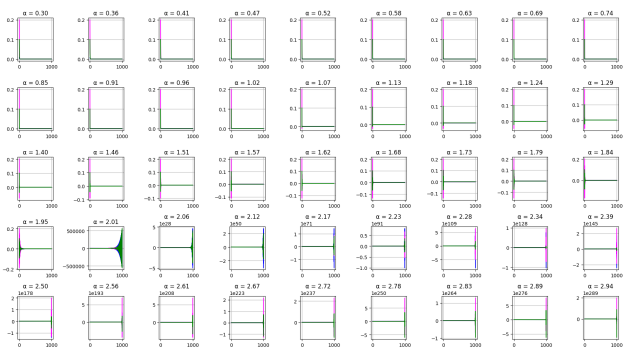

Fig. 4: Calculation result for N=3 by timesteps

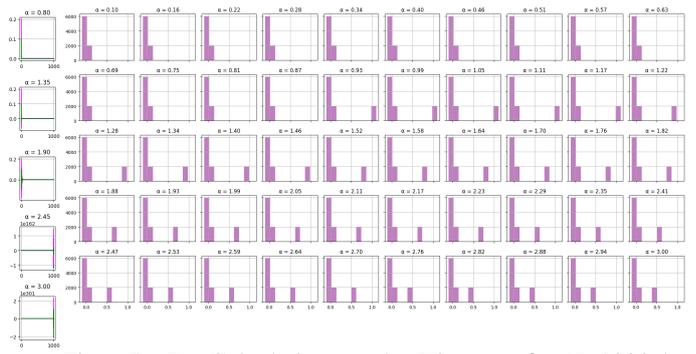

Fig. 5: Pre-Calculation results Histgram for N=2000 by timesteps

(1) **Consensus Formation Among the Three Parties:** From the given plots, it can be observed that the three parties tend to reach a consensus as the value of $\alpha$ increases. Initially, for smaller values of $\alpha$, there is significant divergence in the opinions. However, as $\alpha$ increases, the opinions converge, indicating a tendency towards consensus formation.

(2) **Opinion Formation in the Society:** In real-world scenarios, such plots can be reflective of how opinions form within a society. Initially, there might be a lot of dissenting opinions, but with the right conditions (as represented by $\alpha$ in this context), society can move towards a general consensus. However, it's worth noting that for certain values of $\alpha$, the system can show erratic behavior, which could be likened to societal unrest or instability.

(3) **Insights from Stepwise Variation of $\alpha$:** Upon a stepwise increase in the value of $\alpha$, there seems to be a gradual shift from divergence to convergence among the opinions. However, there are certain critical values of $\alpha$ where the behavior of the system changes dramatically. These points might be of significant interest as they can represent thresholds or tipping points in the consensus-building process.

(4) **Trend of Opinion Clashes with Respect to $\alpha$:** It's evident that the clash or difference in opinions is inversely proportional to the value of $\alpha$. For smaller $\alpha$ values, the differences are more pronounced. But as $\alpha$ increases, the clashes reduce, indicating a harmonization of views.

## 5. Conclusion

In Fig.5, analysis from the Graphs,

**Consensus Formation among 2000 Individuals**

From the graphs, we observe that as the value of $\alpha$ increases, the distribution of opinions becomes more polarized. At low values of $\alpha$, the opinions are somewhat evenly distributed, but as $\alpha$ grows, the opinions cluster around certain values, specifically 0 and 1, which can be considered as extreme opinions.

**Opinion Formation in Society**

In real-world social contexts, these results may reflect the way opinions form under different societal pressures. A low $\alpha$ could represent a society with weak peer influence or with a strong emphasis on individualism, where people's opinions are more diverse. On the other hand, a high $\alpha$ might signify a society with strong peer influence or external pressures causing opinions to polarize.

**Insights from Step-wise Alpha Values**

Analyzing the graphs at each step of $\alpha$ allows us to see the gradual shift in opinion dynamics. Initially, for smaller $\alpha$ values, there's a broad spread of opinions. But as we move to larger $\alpha$ values, there's a clear trend towards opinion polarization. It suggests that the system's sensitivity to the parameter $\alpha$ is non-linear.

**Collisions in Opinions with Alpha Values**

For higher values of $\alpha$, it's evident that the opinions are colliding more frequently towards the extremes. This phenomenon indicates that when external pressures or influence (represented by $\alpha$) is strong, it can lead to more frequent clashes or stronger alignment of opinions, reducing the middle ground. In this research, we presented a theory of opinion dynamics that considers the opinion of each person a continuous value, rather than a discrete value. Opinions are represented by real numbers ranging from positive to negative. We introduce "trust" and "distrust" as a coefficient of each person pairs. In addition to the influence of opinion exchanges within each group, we constructed a mathematical model that incorporates external pressure. Using this theory, we can mathematically express many phenomena that can occur in a group in society.

In this new opinion dynamics theory, it is possible to calculate the dynamics of a complicated system mixed with people's trust and suspicion. Also, as there is no upper limit on the opinion, we can explain the situation where opinions are getting sharper and sharper. Simulation of a large number of people is also prepared. In the future, we will compare and examine which case is assumed whether this theory conforms to actual data concerning speech in actual political and social problems. In the future, we will compare and examine which case is assumed whether this theory conforms to actual data concerning speech in actual political and social problems.

# Aknowlegement

The author is grateful for discussion with Prof. Serge Galam. This research is supported by Grant-in-Aid for Scientific Research Project FY 2019-2021, Research Project/Area No. 19K04881, "Construction of a new theory of opinion dynamics that can describe the real picture of society by introducing trust and distrust".